# Chapter 12: Metadata and provenance management

Ewa Deelman[1], Bruce Berriman[2], Ann Chervenak[1], Oscar Corcho[3], Paul Groth[1], Luc Moreau[4],

[1]USC Information Science Institute, Marina del Rey, CA

[2]Caltech, Pasadena, CA

[3]Universidad Politécnica de Madrid, Madrid, ES

[4]University of Southampton, Southampton, UK

**Abstract**

Scientists today collect, analyze, and generate TeraBytes and PetaBytes of data. These data are often shared and further processed and analyzed among collaborators. In order to facilitate sharing and data interpretations, data need to carry with it metadata about how the data was collected or generated, and provenance information about how the data was processed. This chapter describes metadata and provenance in the context of the data lifecycle. It also gives an overview of the approaches to metadata and provenance management, followed by examples of how applications use metadata and provenance in their scientific processes.

## 1    Metadata and Provenance

Today data are being collected by a vast number of instruments in every discipline of science. In addition to raw data, new products are created every day as a result of processing existing data and running simulations in order to understand observed data. As the sizes of the data sets grow into the petascale range, and as data are being shared among and across scientific communities, the importance of diligently recoding the meaning of data and the way they were produced increases dramatically.

One can think of *metadata* as data descriptions that assign meaning to the data, and data *provenance* as the information about how data was derived.  Both are critical to the ability to interpret a particular data item. Even when the same individual is collecting the data and interpreting them, metadata and provenance are important. However, today, the key drivers for the capture and management of data descriptions are the scientific collaborations that bring collective knowledge and resources to solve a particular problem or explore a research area.  Because sharing data in collaborations is essential, these data need to contain enough information for other members of the collaboration to interpret them and then use them for their own research.  Metadata and provenance information are also important for the automation of scientific analysis where software needs to be able to identify the data sets appropriate for a particular analysis and then annotate new, derived data with metadata and provenance information.

Figure 12.1 depicts a generic data lifecycle in the context of a data processing environment where data are first discovered by the user with the help of metadata and provenance catalogs. Next, the user finds available analyses that can be performed on the data, relying on software component libraries that provide *component metadata*, a logical description of the component capabilities. During the data processing phase, data replica information may be entered in replica catalogs (which contain metadata about the data location), data may be transferred between storage and execution sites, and software components may be staged to the execution sites as well. While data are being processed, provenance information can be automatically captured and then stored in a provenance store. The resulting derived data products (both intermediate and final) can also be stored in an archive, with metadata about them stored in a metadata catalog and location information stored in a replica catalog.

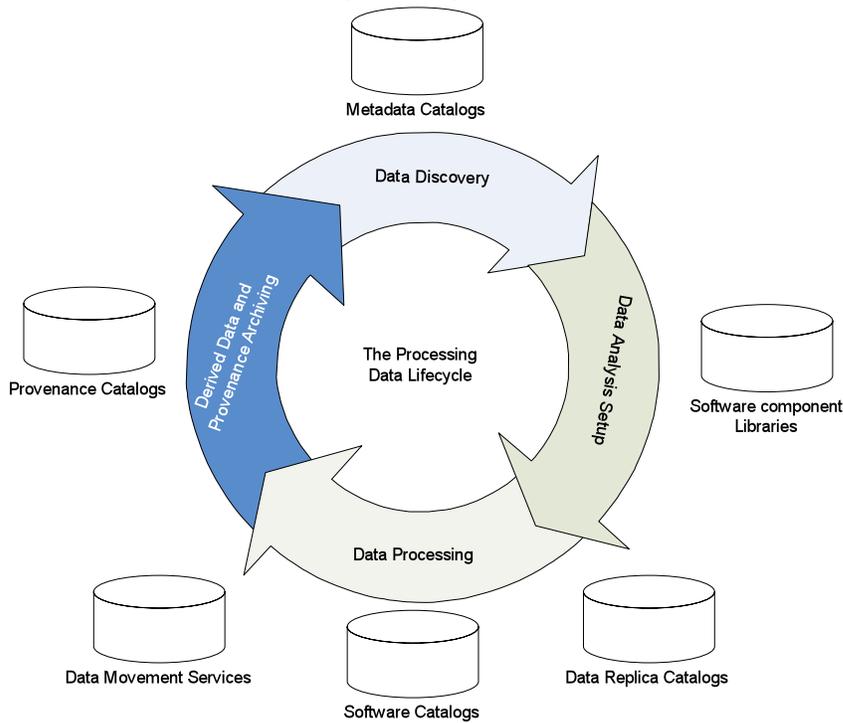

**Figure 12.1: The Data Lifecycle.**

## 2 Metadata

From a general point of view, metadata may be defined as "data about data". However, this definition is too broad; hence other more specific definitions for this term have been provided in the literature, each of them focusing on different aspects of metadata capture, storage and use. Probably one of the most comprehensive definition is the one from [1], which defines metadata as "structured data about an object that supports functions associated with the designated object". The structure implies a systematic data ordering according to a metadata schema specification; the object can be any entity or form for which contextual data can be recorded; and the associated functions can be activities and behaviors of the object. One of the characteristics of this definition is that it covers the dual function that metadata can have: describing the objects from a logical point of view as well as describing their physical and operational attributes.

We can cite a wide range of objects that metadata can be attached to, such as databases, documents, processes and workflows, instruments and other resources. These objects may be available in different formats. For example, documents may be available electronically in the form of HTML, PDF, Latex, etc., in the Web, in a data Grid, on a PC hard disk, or on paper in a library, among other formats. At the same time, metadata can be also expressed in a wide range of languages (from natural to formal ones) and with a wide range of vocabularies (from simple ones, based on a set of agreed keywords, to complex ones, with agreed taxonomies and formal axioms). Metadata can be available in different formats, both electronic and on paper, for example, written in a scientist's lab notebook or in the margins of a textbook. Metadata can also be created and maintained using different types of tools, from text editors to metadata generation tools, either manually or automatically.

Given all this variety in representation formats, described resources, approaches for metadata capture, storage and use, etc., there is not a commonly agreed taxonomy of types of metadata or types of described resources, but different points of view about how metadata can be generated and used. We will now go through some of these points of view, illustrating them with examples.

One of the properties of metadata is that it can be organized in layers, that is, metadata can refer to raw data, (e.g. coming from an instrument or being available in a database), refer to information about the process of obtaining the raw data, or refer to derived data products. This allows distinguishing different layers (or chains) of metadata: primary, secondary, tertiary, etc. As an example, let us consider an application in the satellite imaging domain, such as the one described in [2]. Raw data coming from satellites (e.g., images taken by instruments in the satellite) are sent to the ground stations so that they can be stored and processed. A wide range of metadata can be associated with these data, such as the times when they were obtained and transferred, the instrument used for capturing them, the time period when the image was taken, the position to which it refers, etc. This is considered as the primary metadata of the images received. Later on, this metadata can be used to check whether all the images that were supposed to be obtained from an instrument in a period of time have been actually obtained or whether there are any gaps, and new metadata can be generated regarding the grouping of pieces of metadata for an instrument, the quality of the results obtained for that time period, statistical summaries, etc. This is considered as secondary metadata, since it does not refer to the raw data being described, but to the metadata that refer to the analysis, summaries, and observations about the raw data, so that it forms a set of layers or a chain of metadata descriptions. Another common example of this organization of metadata into layers is that of provenance, which is described in the next section.

In all these cases, it is important to determine which type (layer) of metadata we use for searching, querying, etc., and which type of metadata we show to users, so that metadata coming from different layers is not merged together and is shown with the appropriate level of detail, as discussed in [3] .

The organization of metadata into layers also reflects an interesting characteristic of how metadata is used. To some extent, what is metadata for one application may be considered as data for another. In the previous example in the satellite domain, metadata about the positions of images on the Earth is considered as part of the primary metadata that is captured and stored for the satellite mission application when the information arrives to the ground station. However, the same spatial information would be considered as a data source for other applications, such as a map visualization service (e.g., Google Earth) that positions those resources in a map. In contrast, the dates when the images were taken or the instruments with which they were produced may still be considered as metadata in both cases.

Another aspect of metadata comes into play when new data products are being generated either as a result of analysis or simulation. These derived data are now scientific products and need to be described with appropriate metadata information. In some disciplines such as astronomy (see Section 5.1), the community has developed standard file formats that include metadata information in the header of the file. These are often referred to as "self-describing data formats", since each file stored in such a format has all the necessary metadata in its header. Software is then able to read this metadata and to generate new data products with the appropriate headers. One of the difficulties of this approach is to be able to automatically catalog the derived data. In order to do that some process needs to be able to read the file headers and then extract the information and place it in a metadata catalog. In terms of metadata management, astronomy seems to be ahead of other disciplines, possibly because in addition to the astronomy community, the discipline appeals to many amateurs. As a result, astronomers needed to face the issue of data and metadata publication early on, making their data broadly accessible. Other disciplines of science are still working on the development of metadata standards and data formats. Without those, software cannot generate descriptions of the derived data. Even when the standards are formed within the community, there are often a number of legacy codes that need to be retrofitted (or wrapped) to be able to generate the necessary metadata descriptions as they generate new data products.

Finally, another perspective about metadata is whether a piece of metadata reflects an objective point of view about the resources that are being described or only a subjective point of view about it. While in the former case the term "metadata" is generally used, in the latter the more specific term "annotation" is more commonly used. Annotations are normally produced manually by humans and reflect the point of view of those humans with respect to the objects being described. These annotations are also known as "social annotations", to reflect the fact that they can be provided by a large number of individuals. They normally consist of sets of tags that are manually attached to the resources being described, without a structured schema to be used for this annotation or a controlled vocabulary to be used as a reference. These types of annotations provide an additional point of view over existing data and metadata, reflecting the common views of a community, which can be extracted from the most common tags used to describe a resource. Flickr [4] or del.icio.us [5] are examples of services used to generate this type of metadata for images and bookmarks respectively, and are being used in some cases in scientific domains [3].

There are also other types of annotations that are present in the scientific domain. For example, researchers in genetics annotate the Mouse Genome Database [6] with information about the various genes, sequences, and phenotypes. All annotations in the database are supported with experimental evidence and citations and are curated. The annotations also draw from a standard vocabulary (normally in the form of controlled vocabularies, thesauri or ontologies, as described in the following subsection), so that they can be consistent. Another example of scientific annotations is in the neuroscience domain, where scientists are able to annotate a number of brain images [7]. The issue in brain imaging is that there are very few automated techniques that can extract the features in an image. Rather, the analysis of the image is often done by a scientist. In some cases, the images need to be classified or annotated based on the functional properties of the brain, information which cannot be automatically extracted. As with other annotations, brain images can be annotated by various individuals, using different terms from an overall vocabulary. An advantage of using a controlled vocabulary for the annotations is that the annotations can be queried and thus data can be discovered based on these annotations.

Annotations can also be used in the context of scientific workflows (see chapter 13), where workflow components or entire workflows can be annotated so that they can be more readily discovered and evaluated for suitability. The myGrid project [8] has a particular emphasis on bioinformatics workflows composed of services broadly available to the community. These services are annotated with information about their functionality and characteristics. myGrid annotations can be both in a free text form and drawn from a controlled vocabulary [9].

## 2.1 The Role of Ontologies in Metadata Specification

Together with controlled vocabularies and thesauri, ontologies have become one of the most common means to specify the structure of metadata in scientific applications, such as the previous ones. Ontologies are normally defined as "formal, explicit specifications of shared conceptualizations" [10]. A conceptualization is an abstract model of some phenomenon in the world by having identified the relevant concepts of that phenomenon. Explicit means that the type of concepts used, and the constraints on their use are explicitly defined. Formal refers to the fact that the ontology should be machine-readable. Shared reflects the notion that an ontology captures consensual knowledge, that is, it is not private of some individual, but accepted by a group.

Ontologies started to be used for this purpose in document metadata annotation approaches in pre-Semantic Web applications like the SHOE project [11], the (KA)$^2$ initiative [12], and Planet-Onto [13], among others. Later ontologies have become a commodity for specifying the schema of metadata annotations, not only about Web documents, but also about all types of Web and non-Web resources. The benefits that they provide with respect to other artifacts are mainly related to the fact that they capture consensual knowledge (what facilitates interoperability between applications, although consensus is sometimes difficult to achieve, as described in the previous section), and that they can be used to check the consistency of the annotations and to infer new knowledge from existing annotations, since they are expressed in formal languages with a clear logical theory behind.

Not all ontologies have the same degree of formality; neither do they include all the components that could be expressed with formal languages, such as concept taxonomies, formal axioms, disjoint and exhaustive decompositions of concepts, etc. Given this fact, ontologies are usually classified either as lightweight or heavyweight [10]. An example of the former would be Dublin Core[1], which is being widely used to specify simple characteristics of electronic resources, specifying a predefined set of features such as *creator*, *date*, *contributor*, *description*, *format*, etc. Examples of the latter would be the ontologies used for workflow annotation in the myGrid project or for product description in the aforementioned satellite imaging application. Lightweight ontologies can be specified in simpler formal ontology languages like RDF Schema [14], and heavyweight ontologies require more complex languages like OWL.

## 3 Provenance

Provenance is commonly defined as the origin or source or history of derivation of some object. In the context of art, this term carries a more concrete meaning: it denotes the record of ownership of an art object. In this context, such concrete records allow scholars or collectors to verify and ascertain the origin of the work of art, its authenticity and therefore its price.

This notion of provenance can be transposed to electronic data [15]. If the provenance of data produced by computer systems could be determined as it can for some works of art, then users would be able to understand how documents were assembled, how simulation results were determined, or how analyses were carried out. For scientists, provenance of scientific results would indicate how results were derived, what parameters influenced the derivation, what datasets were used as input to the experiment, etc. In other words, provenance of scientific results would help *reproducibility* [16], a fundamental tenet of the scientific method.

---

[1] http://www.dublincore.org/

Hence, in the context of computer systems, we define *provenance of a data product as the process that led to such a data product*, where process encompasses all the derivations, datasets, parameters, software and hardware components, computational processes, digital or non-digital artifacts that were involved in deriving and influencing the data product.

Conceptually, such provenance could be extremely large, since potentially it could bring us back to the origin of time. In practice, such level of information is not required by end users, since their needs tend to be limited to specific tasks, such as experiment reproducibility or validation of an analysis.

To support the vision of provenance of electronic data, we make the distinction between *process documentation*, a representation of past processes as they occur inside computer systems, and *provenance queries* extracting relevant information from process documentation to support user's needs.

Process documentation is collected during execution of processes or workflows and begins to be accumulated well before data are produced, or even before it is known that some data set is to be produced. Hence, management of such process documentation is different from metadata management. In practice, in a given application context, users may identify commonly asked provenance queries, which can be pre-computed, and for which the results are stored and made available.

Similar to the earlier discussion of different metadata layers, we can think of provenance as consisting of descriptions at different levels of abstraction, essentially aimed at different audiences: to support scientific reproducibility, engineering reproducibility, or even deeper understanding of the process that created the derived data (we provide an example of the latter in the context of scientific workflows below). In terms of scientific reproducibility, where scientists want to share and verify their findings with colleagues inside or outside their collaboration, the user may need to know what data sets were used, and what type of analysis with what parameters were used. However, in cases where the results need to be reproduced bit-by-bit, more detailed information about the hardware architecture of the resource, environment variables used, library versions, etc. are needed. Finally, provenance can also be used to analyze the performance of the analyses [17], where the provenance records are mined to determine the number of tasks executed, their runtime distribution, where the execution took place, etc.

In some cases, scientific processes are managed by workflow management systems. These may take in an abstract workflow description and generate an executable workflow. During the mapping the workflow system may modify the executable workflow to the point that it is no longer easy to map between what has been executed and what the user specified [18]. As a result, information about the workflow restructuring process needs to be recorded as well [19]. This information allows us not only to relate the user-created and the executable workflow but is also the foundation for workflow debugging, where the user can trace how the specification they provided evolved into an executable workflow.

In the area of workflow management and provenance, an interesting aspect of workflow creation is the ability to re-trace how a particular workflow has been designed, or in other words, to determine the provenance of the workflow creation process. A particularly interesting approach is taken in VisTrails [20, 21] where the user is presented with a graphical interface for workflow creation and the system incrementally saves the state of the workflow as it is being designed, modified, or enhanced. As a result the user may re-trace their steps in the design process, choose various "flavors" of the same workflow and try and retry different designs. A challenge could be not only to capture the "how", but also the "why" of the design decisions made by the users.

Unlike metadata, much process documentation is relatively easy to produce automatically, especially in the context of workflows, since the workflow system is in charge of setting up the environment for the computation, managing the data, and invoking the analysis steps. Thus, the workflow system is able to capture the information about where the execution took place, what were the parameters and environment variables used by a software component, which data files where used, etc. Some of that information may be somewhat obscured, for example, when a configuration file is used instead of placing the parameters on the command line. However, the workflow system can also automatically save information about which configuration file was used. It is also interesting to note that the capabilities of workflow management and provenance management systems are complementary (process execution vs. process documentation), and thus it is possible to integrate workflow management systems with provenance management systems that have been developed independently of each other [22].

## 4   Survey of existing approaches

In this section we describe some of the existing approaches for managing metadata and provenance. We start by giving an example of the various types of attributes that are part of the metadata of scientific data.

### *4.1   Metadata Schema and Metadata Attributes*

Metadata attributes that are elements of a metadata schema can encompass a variety of information. Some metadata is application independent, such as the creation time, author, etc. described in Dublin Core [23], while other metadata is application dependent and may include attributes such as the duration of an experiment, temperature of the device, etc. Many applications have expanded the Dublin Core schema to include application-dependent attributes [24].

Based on experiences with a number of scientific applications, we described nine general types of metadata attributes [25, 26]. These descriptions were used in metadata systems developed as part of the Metadata Catalog Service (MCS) [25], the Laser Interferometer Gravitational-Wave Observatory (LIGO) project [27], the Linked Environments for Atmospheric Discovery (LEAD) project [28], and others. Below, we describe some of metadata attribute categories.

*Logical File Metadata*: Metadata attributes associated with the logical file include the following. A *logical file name* attribute specifies a name that is unique within the namespace managed by a metadata service. A *data type* attribute describes the data item type, for example, whether the file format is binary, html, XML, etc. A *valid* attribute indicates whether a data item is currently valid, allowing users to quickly invalidate logical files, for example, if the file administrator determines that a logical file contains incorrect data. If data files are updated over time, a *version* attribute allows us to distinguish among versions of a logical file. A *collection identifier* attribute allows us to associate a logical file with exactly one logical collection. *Creator* and *last modifier* attributes record the identifications of the logical file's creator and last modifier. Other attributes specify the *creation time* and the *last modification time*. A *master copy* attribute can contain the physical location of the definitive or master copy of the file for use by higher level data consistency services.

*Logical collection metadata:* Attributes include the *collection name* and a description of the *collection contents*, which consists of the list of logical files and other logical collections that compose this collection. Each logical file can belong to at most one logical collection. Logical collections may contain other collections, but must form an acyclic collection hierarchy. In addition, the collection metadata includes a *text description* of the collection, information about the *creator* and *modifiers* of the collection, and *audit* information. Finally, there may be a parent attribute that records the identifier of the parent logical collection. There may be an arbitrarily deep acyclic hierarchy of logical collections.

*Logical view metadata:* Attributes include the *logical view name* and *description*; information about the logical files, logical collections and other logical views that compose this logical view; attributes describing the *creator* and *modifiers* of the view, and *audit* information.

*Authorization metadata:* Attributes are used to determine the *access permissions* to the data items and metadata.

*User metadata:* Attributes that describe *writers of metadata*, including contact information. The attributes specify the distinguished *name, description, institution, address, phone* and *email* information for writers.

*User-defined metadata* attributes: Extensibility of the schema beyond predefined attributes is provided by allowing users to define new attributes and associate them with logical files, collections or views. Extensibility is an essential requirement in metadata systems, since each scientific application domain typically produces one or more metadata schemas that capture attributes of interest to that community.

*Annotation* attributes: Annotations can be attached to logical files, collections or views. Annotation metadata includes the *identifier* for the object being annotated and the *object type* (logical file, collection or view). The annotation attribute is a string provided by the user. Annotation metadata also includes the distinguished name of the user creating the annotation and a *timestamp* that records when the annotation was created.

*Creation and transformation history metadata:* These provenance attributes record process information about how a logical file was created and what subsequent transformations were performed on the data. This information may be used to recreate the data item if it ever gets corrupted, or the application may decide to recreate the dataset if the cost of recreating it is less than the cost of retrieval.

*External catalog metadata***:**  Because metadata may be spread across multiple heterogeneous catalogs, these attributes can be used to access external catalogs.

## 4.2    Technologies for storing metadata management

Metadata catalogs have utilized a variety of underlying technologies, including relational databases, XML-based databases, Grid database services and RDF triple stores.

Relational databases are well-suited for metadata repositories in application domains that have a well-defined metadata ontology that changes relatively slowly. Relational databases store data in tables and offer good scalability in both the amount of data they can store and the number of simultaneous queries they can support. These databases also support the construction of indexes on particular metadata attributes, which can provide good performance for common queries related to those attributes. Scientific collaborations often rely on open source relational databases such as PostgreSQL [29] and MySQL [30], but some projects use commercial solutions (Oracle, DB2). Examples of scientific collaborations whose metadata catalogs have used a relational database include the LIGO project [31] and the Earth System Grid [32].

XML-based databases provide the ability to store and query content stored in eXtended Markup Language (XML) format. While some "native" XML databases store data in XML format, others map XML data to a different format and use a relational or hierarchical database to store the data. XML databases can be queried using a variety of languages, such as XPath and XQuery. Examples of XML databases include the Apache Xindice database [33] and Oracle Berkeley DB XML [34].

The Resource Description Framework (RDF) [35] supports the representation of graph-based semantic information using a simple data model. An RDF expression is represented by a set of triples, where each triple contains a subject, predicate and object. A triple asserts that a relationship exists between the subject and the object, where the relationship is specified by the predicate. Information about the predicate, subject and object of the triples may be related to components defined in an existing ontology (which can be implemented in languages like RDF Schema or OWL). This allows defining explicitly the semantics of the objects used in the triples, and of the assertions made within these triples. Besides, it allows performing consistency checks and inferring new information from the information provided in the triples.

RDF information can be stored and queried in an RDF triple store. Over time, a growing number of metadata catalogs have made use of RDF to store semantic metadata information. RDF triple stores are often implemented using a relational database or a hierarchical database as a back end. For example, the Jena semantic web toolkit [36] includes functionality to store and query RDF triples using an Oracle Berkeley DB back end or using a relational database (PostgreSQL, MySQL, etc.) via a JDBC interface [37]. Sesame [38] provides a Java framework for storing and querying RDF triples. The Storage and Inference Layer (SAIL) of Sesame interfaces between RDF functions and the API for various databases, including relational and object-oriented databases and other RDF stores [37]. Besides basic querying, these triple stores also implement consistency checking and inference services that exploit the semantics defined in RDF Schema and OWL ontologies.

Another technology used in Grid environments to deploy metadata catalogs is the OGSA-DAI (Data Access and Integration) service [39]. The OGSA-DAI middleware provides a grid service interface that exposes data resources such as relational or XML databases. Clients of the OGSA-DAI service can store and query metadata in the back end database. One example of a metadata catalog that uses the OGSA-DAI service in its deployment is the Metadata Catalog Service (MCS) [40, 41], which provides a set of generic metadata attributes that can be extended with application-specific attributes (described in sub-section 4.1). MCS is used to store metadata during workflow execution by the Pegasus workflow management system [18, 42], which in turn is used by a variety of scientific applications, including LIGO [43], SCEC [44] and Montage [45].

Finally, some metadata catalogs are integrated into general-purpose data management systems, such as the Storage Resource Broker (SRB) [46]. SRB includes an internal metadata catalog, called MCAT [47]. SRB supports a logical name space that is independent of the physical name space. Logical files in SRB can also be aggregated into collections. SRB provides various authentication mechanisms to access metadata and data within SRB.

## *4.3*     *Technologies for provenance management*

The topic of provenance is the focus of many research communities, including e-Science and grid computing, databases, visualization, digital libraries, web technologies, and operating systems. Two surveys by Bose and Frew [48] and Simmham [49] provide comprehensive overviews of provenance-related concepts, approaches, technologies and implementations. In the recent Provenance Challenge [50], sixteen different systems were used to answer typical Provenance Queries pertaining to a Brain

Atlas data set that was produced by a demonstrator workflow in the context of functional magnetic resonance imaging.

Inspired by the summary of contributions in [50], we present key characteristics of provenance systems. Most provenance systems are embedded inside an execution environment, such as a workflow system or an operating system. In such a context, embedded provenance systems can track all the activities of this execution environment and are capable of providing a description of data produced by such environments. We characterize such systems as *integrated environments*, since they offer multiple functionality, including workflow editing, workflow execution, provenance collection and provenance querying [21, 51, 52]. Integrated environments have some benefits, including usability and seamless integration between the different activities. From a provenance viewpoint, there is close semantic integration between the provenance representation and the workflow model, which allows efficient representation to be adopted [53]. The downside of integrated systems is that the tight coupling of components rarely allows for their substitution or use in combination with other useful technologies; such systems therefore have difficulties interoperating with others, a requirement of many large scale scientific applications.

In contrast to integrated provenance environments, approaches such as PASOA [54, 55] and Karma [56] adopt separate, autonomous provenance stores. As execution proceeds, applications produce process documentation that is recorded in a storage system, usually referred to as a *provenance store*. Such systems give the provenance store an important role, since it offers long-term, persistent, secure storage of process documentation. Provenance of data products can be extracted from provenance stores by issuing queries to them. Over time, provenance stores need to be managed to ensure that process documentation remains accessible and usable in the long term. In particular, PASOA has adopted a provenance model that is independent of the technology used for executing the application. PASOA was demonstrated to operate with multiple workflow technologies, including Pegasus [19], VDL [57] and BPEL [58]. This approach that favors open data models and open interfaces allows the scientist to adopt the technologies of their choice to run applications. However, a common provenance model would allow for past executions to be described in a coherent manner, even when multiple technologies are involved.

All provenance systems rely on some form of database management system to store their data, and RDF and SQL stores were the preferred technologies. Associated query languages are used to express provenance queries, but some systems use query templates and query interfaces that are specifically provenance oriented, helping users to express precisely and easily their provenance questions without having to understand the underpinning schemas adopted by the implementations.

Another differentiator between systems is the *granularity of the data* that a provenance management system uses to keep track of the origins of the data. Again, the coupling of the provenance approach to the execution technology can influence the capability of the provenance management system from a granularity viewpoint. For instance, some workflow systems that allow for files to be manipulated by command-line programs such as Pegasus tend to track the provenance of files (and not the data they contain). This capability is sufficient in some cases, but is too coarse-grained in others. Systems such as Kepler [51], on the other hand, have specific capabilities to track the provenance of collections. Other systems are capable of tracking the origins of programs, such as VisTrails. The PASOA system has been demonstrated to capture provenance for data at multiple levels of granularity (files, file contents, collections, etc), and its integration with Pegasus showed it could be used to track the change in the workflow produced by the Pegasus workflow compiler.

Systems such as ES3 [59] and PASS [60] capture events at the level of the operating system, typically reconstructing a provenance representation of files. In such a context, workflow scripts are seen as files whose origin can also be tracked.

The database community has also investigated the concept of provenance. Reusing the terminology introduced in this section, their solutions can generally be regarded as integrated with databases themselves: given a data product stored in a database, they track the origin of data derivations produced by views and queries [61]. From a granularity viewpoint, provenance attributes can be applied to tables, rows and even cells. To accommodate activities taking place outside databases, provenance models that support copy and paste operations across databases have also been proposed [62]. Such provenance models begin to resemble those for workflows, and research is required to integrate them smoothly.

Internally, provenance systems capture an explicit representation of the flow of data within applications, and the associated processes that are executed. At some level of abstraction all systems in the recent provenance challenge [63] use some graph structure to express all dependencies between data and processes. Such graphs are directed acyclic graphs that indicate from which ancestors processes and data products are derived. Given such a consensus, a specification for an Open Provenance Model is emerging [64], and could potentially become the lingua franca by which provenance systems could exchange information. We illustrate this model over a concrete example in Section 6.

## 5 Metadata in Scientific Applications

In this section we present a couple of examples of how scientific applications manage their metadata.

### *5.1 Astronomy*

Astronomy has used, for over 25 years, the *Flexible Image Transport System (FITS)* standard [65] for platform-independent data interchange, archival storage of images and spectra, and all associated metadata. It is endorsed by the U.S. National Aeronautics and Space Administration (NASA) and the International Astronomical Union. By design, it is flexible and extensible and accommodates observations made from telescopes on the ground and from sensors aboard spacecrafts. Briefly, a FITS data file is composed of a fixed logical record length of 2880 bytes. The file can contain an unlimited number of header records, 80 bytes long, having a 'keyword=value' format and written as ASCII strings. These headers describe the organization of the binary data and the format of the contents. The headers are followed by the data themselves, which are represented as binary records. The headers record all the metadata describing the science data. Figure 12.2 depicts an instance of this for the metadata of an image of the galaxy NGC 5584 measured by the Two-Micron All Sky Survey (2MASS) [66].

The complete metadata specification of an astronomical observation includes obvious quantities such as the time of the observation, its position and footprint on the sky and the instrument used to make the observation, but also includes much information custom to particular data set. FITS therefore was designed to allow astronomers to define keywords as needed. Nevertheless, FITS has predefined reserved keywords to describe metadata common to many observations. For example, the relationship between the pixel coordinates in an image and physical units is defined by the World Coordinate System (WCS) [65], which defines how celestial coordinates and projections are represented in the FITS format as keyword=value pairs. These keywords are listed in sequence in Figure 12.2: they start at CTYPE and end at CDELT.

Tabular data and associated metadata are often represented in FITS format, but the FITS standard is poorly specified for tabular data. Instead, tabular material, whether they are catalogs of sources or catalogs of metadata, are generally stored in relational databases to support efficient searches. Transfer of these data is generally in the form of ASCII files, but XML formats such as VOTable [67] are growing in use. Metadata describing catalog data can be grouped into three types: semantics, which describe science content (units, standards, etc.); logistical, which describe the structure of the table (e.g. data types, representation of null values, etc.) and statistical, which summarize the contents (number of sources, ranges of data in each column, etc.). The absence of standard specifications for columns has complicated and confused data discovery and access to tabular data. Work originating at the Centre de Donnees astronomiques de Strasbourg (CDS) [68] has been embraced by the International Virtual Observatory Alliance (IVOA) [69] as part of an international effort to rectify this problem. When this work is complete, all column descriptors will have a well defined meaning connected to a hierarchical data model.

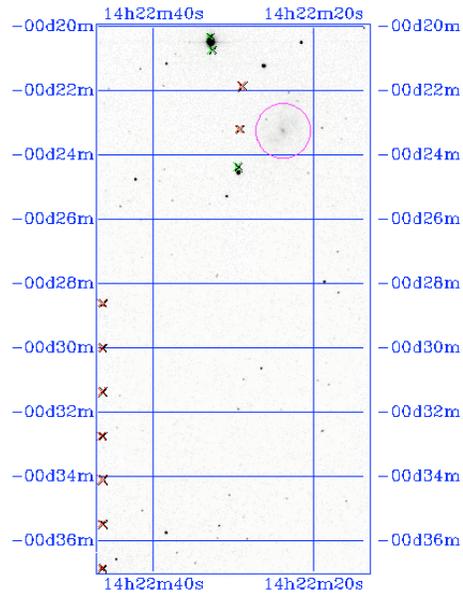

```
ORDATE  = '000503 '           / Observation Ref Date (yymmdd)
DAYNUM  = '1160   '           / Observation Day Num
FN_PRFX = 'j1160059'          / .rdo and .par filename prefix
TYPE    = 'sci    '           / Scan type: dar flt sci cal tst
SCANNO  =              59 /   Scan Number
SCANDIR = 'n     '            / Scan Direction: n, s, -
COMMENT                 (OV)
STRIP_ID=          301788 /   Strip ID (OV)
POSITNID= 's001422 '          / Position ID (OV)
ORIGIN  = '2MASS  '           / 2MASS Survey Camera
CTYPE1  = 'RA---SIN'          / Orthographic Projection
CTYPE2  = 'DEC--SIN'          / Orthographic Projection
CRPIX1  =           256.5 /   Axis 1 Reference Pixel
CRPIX2  =           512.5 /   Axis 2 Reference Pixel
CRVAL1  =      215.6251831 /  RA  at Frame Center, J2000 (deg)
CRVAL2  =     -0.4748106667 / Dec at Frame Center, J2000 (deg)
CROTA2  =    1.900065243E-05 / Image Twist +AXIS2 W of N, J2000 (deg)
CDELT1  =   -0.0002777777845 / Axis 1 Pixel Size (degs)
CDELT2  =    0.0002777777845 / Axis 2 Pixel Size (degs)
USXREF  =           -256.5 /  U-scan X at Grid (0,0)
USYREF  =           19556. /  U-scan Y at Grid (0,0)
```

**Figure 12.2:** Example of Image Metadata in astronomy. On the top is an image of the galaxy NGC 5584, shown at the center of the purple circle. The image was measured as part of the Two Micron All Sky Survey (2MASS). The crosses locate the positions of artifacts in the image. At bottom is a sample of the metadata describing the image, written in the form of keyword=value pairs, in compliance with the definition of the Flexible Image Transport System (FITS) in universal use in astronomy.

## 5.2    Climate Modeling

The Earth System Grid project [32, 70] provides infrastructure to support the next generation of climate modeling research. The Earth System Grid (ESG) allows climate scientists to discover and access important climate modeling data sets, including the Parallel Climate Model (PCM), the Community Climate System Model (CCSM), as well as data sets from the Intergovernmental Panel on Climate Change (IPCC) 4th Assessment Report (AR4) [71].

The original infrastructure for ESG included two data portals, one at the National Center for Atmospheric Research and one at Lawrence Livermore National Laboratory. Each of these ESG portals had an associated metadata catalog that was implemented using a relational database. The next generation of the Earth System Grid features a federated architecture, with data nodes at many sites around the world publishing data through several ESG data portals or *gateways*. This data publication will include the extraction and publication of metadata attributes. The bulk of the metadata for this version of the ESG architecture will be stored in a relational database with a different schema than in the previous generation of ESG. Metadata for research and discovery will also be harvested into an RDF triple store from the multiple federated ESG gateway sites; this will allow users to execute global searches on the full ESG holdings via the triple store.

The metadata model for the latest generation of the Earth System Grid includes metadata classes that describe a climate Experiment, Model, Horizontal Grid, Standard Name, and Project as well as a set of data-related classes [72]. The *Experiment* class includes a specification of the input conditions of a climate model experiment. The *Model* class describes the configuration of a numerical climate model.

The *Horizontal Grid* class specifies a discretization of the earth's surface that is used in climate models. *Standard names* describe scientific quantities or parameters generated by a climate model run, such as air pressure, atmospheric water content, direction of sea ice velocity, etc. A *Project* is an organizational activity that generates data sets. Data objects in the ESG metadata model have one of four types: dataset, file, variable or aggregation. A *dataset* is a collection of data generated by some activity, such as a project, a simulation run, or a collection of runs. *Files* are usually in the self-describing netCDF [73] data format. A *variable* is a data array with n dimensions, where the array is associated with a dataset. An *aggregation* is a service that provides a view of a dataset as a single netCDF file and that can perform statistical summaries over variable values, such as "monthly means".

There are several ways that ESG users search for and select data sets based on metadata attributes [74]. In one scenario, users can search for data sets based on metadata attributes using a simple Google-style text search over all the metadata associated with ESG data sets. The ESG gateway also presents users with a set of search terms and possible values, as illustrated in Figure 12.3; these terms include the experiment, model, domain, grid, variables, temporal frequency, and dataset. An ESG user may also access an ESG portal using a visualization or analysis tool that provides an API with search capabilities that may include issuing metadata queries.

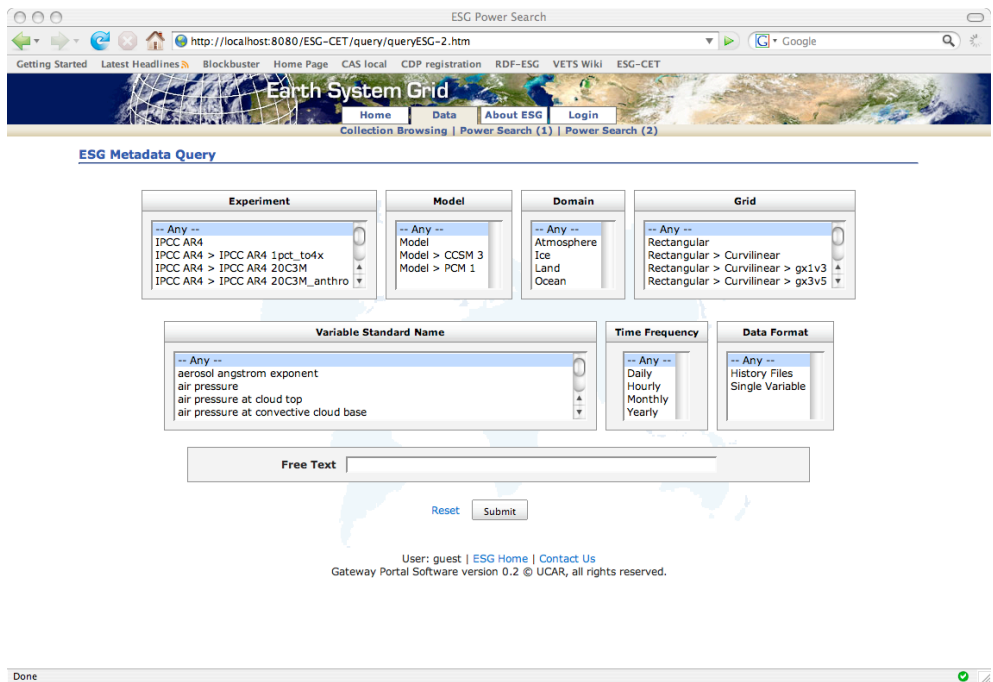

Figure 12.3: Earth System Grid project's interface for metadata queries

# 6    Provenance in Scientific Applications

In this section, we detail how scientists use provenance. First, we discuss the technologies that scientists use everyday in order to determine provenance. We then look at more advanced systems designed for particular applications where provenance plays a critical role; in particular, we focus on geospatial, oceanographic and astronomy applications. Finally, we discuss how open data models will facilitate provenance queries in multi-disciplinary, multi-scale scientific applications.

## *6.1    Provenance in everyday science*

Lab notebooks are used by scientists everyday to record their experimental processes and results. These lab notebooks contain, what we termed previously, process documentation. However, determining provenance from physical notebooks is a laborious procedure. Additionally, lab notebooks are ill equipped to capture the data produced by scientific applications. To ameliorate these concerns, scientists have begun to capture their work using electronic systems. While research on electronic notebooks abounds [75, 76] probably the most widely used system for recording lab notebook style information electronically are Wikis[2]. Wikis provide a system for easily creating, modifying, and collaborating on web pages. For example, the Bradley Laboratory and Drexel University posts the protocols and results for the chemical solubility experiments that they are performing on their Wiki [77]. Similarly, the OpenWetWare project [78] provides a Wiki for the sharing and dissemination of biological protocols. Available protocols range from definitions of how to extract DNA from mouse tissue to setting up microarrays. The project has over 3000 registered users [79].

A key piece of functionality is a Wiki's support for a revision history of each page [80]. This revision history provides a coarse view of the provenance of a page. It provides information as to who edited the page, the time when the page was modified, and the difference between the current state of the page and its last revision. For OpenWetWare, the revision history enables the tracking of a protocol's development and thus allows for the creation of community protocols. Indeed, the ability to roughly determine the provenance of a Wiki page is a key enabler to Open Access Science where scientists share drafts of experiments, papers, and preliminary results online for others to comment on and use [79].

While a Wiki page's coarse provenance is useful, the revision history fails to provide a comprehensive view of the provenance of a page because it does not describe the process by which it was generated. Instead only the difference between pages is known. For example, if a JPEG image added to a Wiki page was produced by converting an image output from an image registration algorithm applied to two other files, the revision history would be unable to inform a scientist that JPEG conversion and image registration were involved in the creation of the JPEG image. Thus, the provenance for the page is incomplete. In many scientific applications, this sort of processing history is required and hence the provenance technologies discussed previously are needed. We now discuss some of these applications.

## *6.2    Provenance in Geospatial Applications*

Some of the first research in provenance was for geospatial systems in particular Geographic Information Systems (GIS) [81]. Knowing the provenance of map products is critical in GIS applications because it allows one to determine the quality of those derived map products. In particular, [82] highlights the need for systems to be able to isolate provenance of a specific spatial region on a map. For example, when computing annual rainfall for the Denver area, the data used as input is the national daily rainfall numbers. When retrieving the provenance of the annual rainfall, it is necessary to have spatial knowledge to know that only the daily rainfall from Denver is used from the input data set.

---

[2] The Oxford English Dictionary defines Wiki as a type of web page designed so that its content can be edited by anyone who accesses it, using a simplified markup language.

A more complex example is given in [83] for tracking the processing of satellite images: Image data covering the western United States and Pacific Ocean is retrieved from National Oceanic and Atmospheric Administration (NOAA) satellites and is sent to a University of California Santa Barbara operated TeraScan ground station. These Advanced Very-High Resolution Radiometer images are processed at Institute for Computational Earth System Science into two data products for southern California and the southwestern United-States: a sea surface temperature map and a near-infrared albedo map. Figure 12.4 shows one such sea surface temperature map. To generate these data products some standard processing steps are applied including calibration, navigation and registration. These data products are then provided to other scientists, who use it for their own process [83].

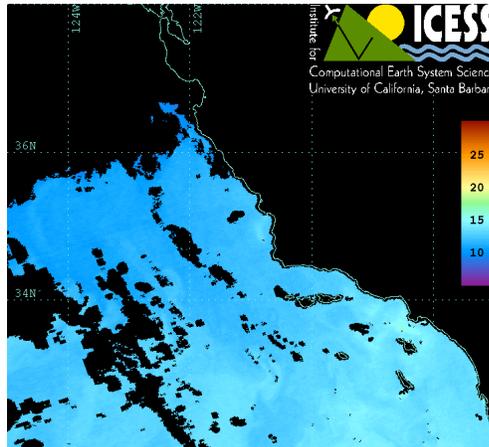

**Figure 12.4: Map of Sea Surface Temperature taken 02/18/2009 2042 GMT**

One of the key observations of [83] is that scientists may use many different command line applications (e.g. Perl scripts, Matlab, TerraScan tools) to generate new satellite data products. The The Earth System Science Workbench (ESSW) system developed at the University of California, Santa Barbara, therefore keeps track of which command-line applications are used to modify the data products and records all intermediate data products. Using ESSW, scientists can determine which satellites were used in generating a data product, what version of the calibration algorithm was used, and the number of scan samples used in the original sensor data.

Provenance is also critical in regenerating large-scale data products derived from satellite imagery. An example is how imagery from the Moderate Resolution Imaging Spectroradiometer Satellite is processed and stored at NASA's Goddard Space Flight Center using the MODIS Adaptive Data Processing System (MODAPS). The imagery obtained directly from the satellite is known as Level 0 data. This raw data is irreplaceable and must be archived. However, the initial transformation of this raw data into more useful data called Level 1B data is too large to be archived. Level 1B data includes calibrated data and geo-located radiances for the 36 channels of the satellite[3]. Level 1B data is then processed and consolidated to create Level 2 data products, which are finally translated into visual images. The Level 2 and final images are the most useful for scientists. Because of its size, Level 1B data is typically kept for 30-60 days and then discarded. To enable this data to be reproduced the MODAPS systems maintains enough process documentation, to reproduce the Level 1B data from the raw satellite data. The process documentation includes the algorithms used, their versions, the original source code, a complete description of the processing environment and even the algorithm design documents themselves. Essentially, provenance enables a *virtual archive* of satellite data, which otherwise would be lost or difficult to recreate [84].

Provenance for geospatial data is also extremely important for merging data from multiple sources. This aspect was discussed in detail in Chapter 10 in the context of interoperability and data integration in geosciences.

## 6.3    Provenance for Oceanographic Applications

When studying the oceans, scientists require data from multiple data sources whether they are shipboard instruments, buoy based sensors, underwater vehicles and permanent stations. These data sources measure everything from the temperature of the ocean to its salinity and the amount of chlorophyll present. These data sources are combined with published data including satellite imagery in simulations to predict oceanographic features, such as seasonal variations in water levels [85]. Using the current data analysis routines, it is difficult to ascertain how the results of these simulations were produced because the information is spread in log files, scripts and notes [85].

To address this problem, the Monterey Bay Aquarium Research Institute has developed their own in-house system, the Shore Side Data System (SSDS), for tracking the provenance of their data products [86]. These data products range from lists of deployed buoys to time series plots of buoy movement. Using SSDS, scientists can access the underlying sensor data but most importantly they can track back from derived data products to the metadata of the sensors including their physical location, instrument and platform. A key part of the system is the ability to automatically populate metadata fields. For example, by understanding that the position of an instrument is caused by the fact that it is located on a mooring platform, the system can traverse the provenance graph to fill in the position metadata for that instrument [86]. It is interesting to note that the metadata produced by SSDS is in the netCDF standard format, which was previously discussed in Section 12.4. SSDS is an example of a production provenance system as it has been used daily for the past four years for managing ocean observation data [86].

The focus of SSDS is tracking the provenance of data sets back to the instruments and the associated configuration metadata. A more complex example of provenance in oceanography is given in [85]. In this work, the authors present a system that combines sensor data products with simulations to present 3D visualizations of fishery data. Specifically, for the Collaborative Research on Oregon Ocean Salmon Project, they combined data about the location and depth of where salmon were caught in the Northwest of the United States with simulation data about ocean currents to generate visualizations of the depth and distribution of fish when looking at the continental shelf [85]. The key use of provenance here is to enable the scientists to explore the parameter space of a visualization without having to worry about tracking the changes to their visualization pipeline. For example, to see a different perspective on the fish, the scientist may have to reconfigure the model they are using. With the VisTrails [87] system, (described briefly in Section 13.5)  they can easily find the changes they made or go back to other visualization pipelines. This functionality is critical when dealing with these complex oceanographic applications that integrate a variety of simulation techniques and data sources.

---

[3] http://modis.gsfc.nasa.gov/data/dataprod/dataproducts.php?MOD_NUMBER=02

## 6.4  End-to-End Provenance for a Large-scale Astronomy Applications

We have seen the need to use provenance to recreate data on demand for satellite imagery, automatically populate metadata fields for oceanographic data, and track the changes in pipelines for visualizing salmon catches. In this section, we see how provenance enables connecting research results to high-level workflows in an astronomy application.

The application we look is Montage [45]. Montage produces science-grade mosaics of the sky on demand. This application can be structured as a workflow that takes a number of images, projects them, adjusts their backgrounds, and adds the images together. A mosaic of 6 degrees square would involve processing 1,444 input images, require 8,586 computational steps and generate 22,850 intermediate data products. Executing the Montage workflow requires potentially numerous distributed resources that may be shared by other users. Because of the complexity of the workflow and the fact that resources often change or fail, it is infeasible for users to define a workflow that is directly executable over these resources. Instead, scientists use "workflow compilers" such as Pegasus [18, 42] (See Chapter 13) to generate the executable workflow based on a high-level, resource-independent description of the end-to-end computation (an *abstract workflow*). This approach gives scientists a computation description that is portable across execution platforms and can be mapped to any number of resources. However, the additional workflow mapping also increases the gap between what the user defines and what is actually executed by the system and thus complicates the interpretation of the results: the connection between the scientific results and the original experiment is lost.

To reconnect the scientific results with the experiment, [19, 88] presents a system for tracking the provenance of a mosaic back to the abstract workflow that it was generated from. The system integrates the PASOA [54] and Pegasus systems to answer provenance questions such as what particular input images were retrieved from a specific archive, whether parameters for the re-projections were set correctly, what execution platforms were used, and whether those platforms included processors with a known floating point processing error.

To accomplish this, each stage of the compilation from abstract workflow to executable workflow is tracked in Pegasus. For example, one of Pegasus's features is to select which sites or platforms each computational step should be executed at. During Pegasus compilation process, this information is stored as process documentation within PASOA's provenance store. Additionally, this information is linked to the subsequent compilation steps such as intermediate data registration, and task clustering. Finally, during the execution of the Pegasus produced workflow, all execution information is stored and linked to the workflow within the provenance store. Using this process documentation, a provenance graph of the resulting sky mosaic can be generated that leads back to the specific site selected.

The availability of provenance in Montage enables astronomers to take advantage of workflow automation technologies while still retain all the necessary information to reproduce and verify their results. Outside of Montage, provenance is an underpinning technology that allows for workflow automation, a technology necessary for other large-scale Grid based science applications, such as astrophysics [89].

## 6.5 Enabling Multi-disciplinary and Multi-scale Applications using Provenance

As we have seen, provenance plays an important role in enabling scientific applications. In particular, those applications that use a variety of heterogeneous data sources or computational resources benefit. Scientific problems are increasingly becoming multi-disciplinary and multi-scale. For example, biomedical applications may combine the results of chemistry simulations of molecular interactions with data about tissues and other organs [90].

To allow the provenance of data to be determined across boundaries of scale, discipline and technologies, there is a need for an interoperability layer between systems. One proposed interoperability specification is the Open Provenance Model [55]. This model provides a good outline of what the community developing provenance technologies believes are the core constituents of a provenance graph. We thus use a graphical representation drawn from the Open Provenance Model to illustrate a concrete provenance graph, as shown in Figure 12.5.

In the representation we adopt, nodes of the graph consist of two entities: artifacts and processes. In the context of scientific workflows as considered here, artifacts are immutable pieces of data, whereas processes are transformations that produce and consume artifacts. Artifacts are represented as circles, and processes are denoted by boxes.

Nodes can be connected by edges expressing causal dependencies between artifacts and processes. The origin of an edge represents an effect, whereas its destination represents a cause: the presence of an edge makes explicit the causal dependency between the effect and its cause. In this presentation, we focus on two types of edges: "wasGeneratedBy" and "used". A "wasGeneratedBy" edge expresses how an artifact was dependent on a process for its generation, whereas a "used" edge indicates that a process relied on some artifacts to be able to complete. An artifact can only be generated by a single process, but it can be used by any number of processes, whereas a process can use and generate any number of artifacts. To be able to distinguish the multiple dependent artifacts a process may rely upon, a notion of role is introduced, allowing the nature of the causal dependency to be characterized explicitly.

Using the above notation, we show a provenance graph generated from the workflow adopted by the Provenance Challenge [63], which is inspired by functional MRI (fMRI) workflows to create population-based "brain atlases" from the fMRI Data Center's archive of high resolution anatomical data [91]. In summary, this workflow produces average images along the axes X, Y, and Z, after aligning each input sample with a reference image. Note that like the other applications discussed, neuroscience applications require provenance [92].

Figure 12.5 illustrates a subset of the provenance graph that is constructed as the Provenance Challenge workflow. Such a graph is best read from right to left: the right identifies an artifact, the Atlas X graphic, representing an averaged image along the X axis; all the causal dependencies that led it to be produced appear to its left. Provenance graphs are directed and acyclic, which means that an artifact or a process cannot be (transitively) caused by itself.

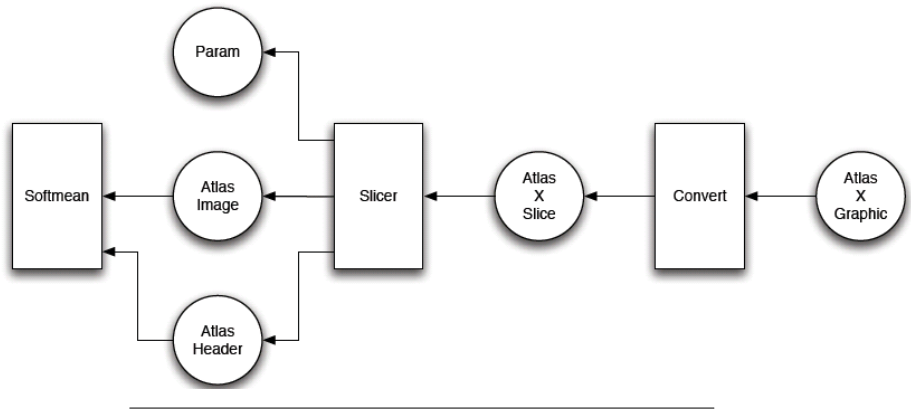

LEGEND:

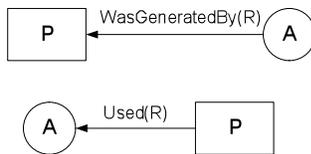

**Figure 12.5: Provenance Graph for the Provenance Challenge Workflow.**

Whenever a scientific workflow system executes this fMRI workflow, it would incrementally produce the various elements of that graph (or an equivalent representation), and store them in a repository, usually referred to as provenance store or provenance catalog. Provenance queries can then be issued to extract a subset of the documentation produced, according to the user's needs.

In conclusion, provenance is critical in many scientific applications ranging from neuroscience to astronomy. As scientific applications become increasingly open and integrated across areas, provenance interoperability becomes an important requirement for systems technologies.

# 7     Current and Future Challenges

There a several challenges in the area of metadata and provenance management. They stem mostly from two facts: 1) scientists need to share information about data within their collaborations and with outside colleagues, and 2) the amount of data and related information is growing at unprecedented scales.

As a result of the scale, users need to decide which data to keep (for example, in high-energy physics only selected and already pre-processed collision events are cataloged). When storing provenance, decisions of what to store need to be made as well. Because it is often hard to predict what will be needed in the future, sometimes data and related information are irrevocably lost.

Because of the size of the collaborations and data sets, data, metadata and provenance information are often not stored at the same location, within the same system. Thus issues of information federation arise.

In some sense, the issue for provenance is not as severe as for metadata. Provenance is in some sense inherently distributed, with information about the data coming from different sources, and it also has explicit links (such as those in the provenance graph) that allow one to follow the provenance trail. Additionally, once the process documentation of an item is generated, it will most likely not change since it is a historical record. On the other hand, metadata about data items may change with time, or a piece of data may be found invalid. As a result, metadata requires more effort in the area of consistency management.

The need to share data results in many challenges. First, communities need to agree on metadata standards. Then, these standards need to be followed by data publishers and software systems so that a consistent view of metadata is maintained. When data are shared across communities, mediation between metadata schemas needs to be performed. The challenge for cross-project or cross-community interoperability is not only technical but also social. How does one motivate scientists to provide the necessary metadata about the primary and derived data? What is the incentive to retrofit the codes and publish the data into community repositories?

In general, future work should focus on extensible metadata and provenance systems that follow common standards are independent of the systems that use them, and can be shared across distributed collaborations. Such systems should support common languages for responding to provenance queries. There is already good progress, but unified metadata and provenance systems for scientific communities are a long way off.

## Acknowledgments

Ewa Deelman's work was funded by the National Science Foundation under Cooperative Agreement OCI-0438712 and grant # CCF-0725332.

Bruce Berriman is supported by the NASA Multi Mission Archive and by the NASA Exoplanet Science Institute at the Infrared Processing and Analysis Center, operated by the California Institute of Technology in coordination with the Jet Propulsion Laboratory (JPL).

Oscar Chorcho's work was funded by the SemsorGrid4Env project (FP7-ICT-223913).

The authors would like to thank members of the Earth System Grid for the use of their metadata example. The Earth System Grid Center for Enabling Technologies is funded by the DOE SciDAC program.